\documentclass[english,aps,prl,twocolumn]{revtex4-1}
\usepackage{bm}
\usepackage{amsbsy}
\usepackage{amssymb}
\usepackage{graphicx}
\usepackage{esint}
\usepackage{color}
\usepackage{array}
\usepackage{textcomp}
\usepackage{multirow}
\usepackage{upgreek}
\usepackage{braket}
\usepackage{babel}
\usepackage{hyperref}
\hypersetup{colorlinks=true, citecolor=blue, urlcolor=blue, linkcolor=blue}

\begin{document}
\title{Long-lasting Molecular Orientation Induced by a Single Terahertz Pulse}

\author{Long Xu}
\thanks{L. X. and I. T. contributed equally to this work.}
\author{Ilia Tutunnikov}
\thanks{L. X. and I. T. contributed equally to this work.}
\author{Erez Gershnabel}
\author{Yehiam Prior}
\author{Ilya Sh. Averbukh}
\email{ilya.averbukh@weizmann.ac.il}
\affiliation{AMOS and Department of Chemical and Biological Physics, The Weizmann
Institute of Science, Rehovot 7610001, Israel}

\begin{abstract}


Control of molecular rotations by laser fields is an active area of research focusing on the alignment/orientation of otherwise isotropic molecular samples. Oriented molecules are useful in many applications, such as molecular orbital tomography, femtosecond imaging of molecular structure, dynamics and chemical reaction control.
In this work, we present a novel, previously unreported phenomenon
appearing in thermal ensembles of non-linear polar molecules excited
by a single THz pulse.
We find that the induced orientation lasts long after the excitation pulse is over. In the case of symmetric-top molecules, the ensemble averaged orientation remains indefinitely constant after the pulse, whereas in the case of asymmetric-top molecules the orientation persists for a long time after the end of the pulse. We discuss the underlying mechanism, study its non-monotonous temperature dependence, and show that there exists an optimal temperature for the maximal residual orientation. The persistent orientation implies long lasting macroscopic dipole moment, which may be probed by second (or higher order) harmonic generation.

\end{abstract}
\maketitle

\emph{Introduction}.--- Molecular alignment and orientation, mostly
by ultrashort optical pulses, have attracted considerable interest
owing to their extreme importance in a variety of applications in
physics, chemical reaction dynamics, and attosecond electron dynamics
(for reviews see, e.g. \citep{Stapelfeldt2003Colloquium,Yasuhiro2010Coherent,Fleischer2012Molecular,Mikhail2013Manipulation,Koch2019Quantum}
and references therein). Independently, rapid advances in terahertz
(THz) technology introduced high energy tabletop sources of ultrashort
THz pulses \citep{Yeh2007}. In recent years, THz pulses were utilized
to orient polar linear molecules \citep{Fleischer2011Molecular,Kitano2013Orientation,Egodapitiya2014Terahertz},
and later on, symmetric- \citep{Babilotte2016Observation} and asymmetric-top
molecules \citep{Damari2016Rotational}.
The pioneering theoretical studies of the hybrid route to orientation, combining THz and optical pulses \cite{Osman2003Evolutionary, Daems2005Efficient,Gershnabel2006Orientation},
were followed by extended theoretical research \cite{Kitano2011High, Shu2013Field, YOSHIDA2015Control, Mirahmadi2018Dynamics} inspired by the recent development of the THz technology, and finally resulted in an experimental demonstration of enhanced orientation \cite{Egodapitiya2014Terahertz}.
The free motion of symmetric-
and asymmetric-tops is a standard textbook, classical and quantum
mechanical topic, but the addition of interaction with an external
field complicates the problem considerably, with just a few known
integrable special cases (see \citep{Arango2008,Schatz2018Symmetric} and references therein).
When a linear polar molecules is excited by a THz pulse, the induced
orientation shows a short transient signal riding on zero baseline,
with periodic recurrences (quantum revivals \citep{Felker1992Rotational,Averbukh1989Fractional,Robinett2004Quantum}),
while keeping the time averaged signal zero. Clearly, the alignment
and orientation of thermal molecular gases require ensemble averaging,
bringing yet another level of complexity.

In this work, we theoretically study, classically and quantum-mechanically, the orientation of a rarified gas of symmetric- and asymmetric-top molecules excited by a single linearly polarized THz pulse.
We demonstrate that the THz
pulse not only induces a short time transient dipole signal (transient
orientation), but also results in a steady molecular orientation that
persists for a long time after the end of the pulse. In the case
of symmetric-top molecules, the ensemble-averaged long-time orientation
remains constant, while for asymmetric-top molecules, it persists
on a time scale exceeding the duration of the pulse by several orders
of magnitude. This behaviour is characteristic for complex molecules,
and it is absent in the case of commonly studied linear molecules.
We show that this phenomenon is classical in nature and its underlying
mechanism is revealed and discussed. The degree of orientation has
a non-monotonous temperature dependence, which is demonstrated and
qualitatively explained.

\begin{figure}[!t]
\centering{}\includegraphics[width=1\linewidth]{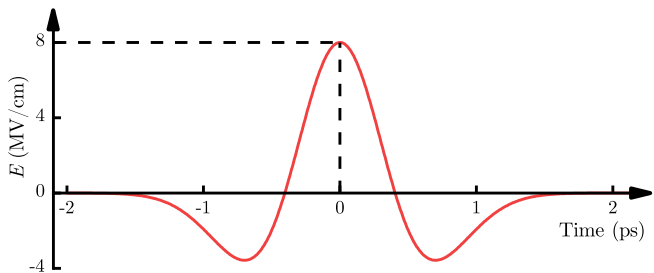} \caption{Time-dependent field amplitude of the THz pulse defined by $E(t)=E_{0}(1-2\kappa t^{2})e^{-\kappa t^{2}}$,
where $E_{0}=8.0\;\mathrm{MV/cm}$ and $\kappa=3.06\;\mathrm{ps^{-2}}$.
Note that the time integral of the electric field is identically zero.
\label{fig:THz-field}}
\end{figure}

\emph{Equations of motion}.---The Hamiltonian describing molecular
rotation driven by external time-dependent field interacting with
the molecular dipole is given by \citep{Krems2018Molecules} $H(t)=H_{R}+H_{\mathrm{int}}(t),$
where $H_{R}$ is the rotational kinetic energy Hamiltonian and $H_{\mathrm{int}}(t)=-\bm{\upmu}\cdot\mathbf{E}(t)$
is the molecular dipole-field interaction. The electric field of the
THz pulse is modeled as $\mathbf{E}(t)=E_{0}(1-2\kappa t^{2})e^{-\kappa t^{2}}\mathbf{e}_{Z}$
\citep{Coudert2017Optimal}. Here $E_{0}$ is the peak amplitude,
$\kappa$ determines the width of the pulse, and $\mathbf{e}_{Z}$
is the unit vector along the laboratory $Z$ axis. Note that the time
integral of the electric field is zero. Figure \ref{fig:THz-field}
shows the shape of the field, for parameters typical for currently
available THz pulses \citep{Clerici2013Wavelength,Oh2014Generation,Vicario2014Infrared,Shalaby2015Demonstration}.
Here, we consider the problem both classically and quantum mechanically.
For the quantum mechanical treatment, the wave function is expanded
in the basis of free symmetric-top wave functions $|JKM\rangle$ \citep{zare1988Angular}.
Here $J$ is the total angular momentum, while $K$ and $M$ denote
its projections on the molecule-fixed axis and the laboratory-fixed
$Z$ axis, respectively. The time-dependent Schrödinger equation $i\hbar\partial_{t}|\Psi(t)\rangle=H|\Psi(t)\rangle$
is solved by numerical exponentiation of the Hamiltonian matrix (see
Expokit \citep{sidje1998Expokit}). We average the orientation signal
over the initial thermal state of the molecular rotor. In the case
of symmetric-top molecules, the nuclear spin statistics \citep{McDowell1990Rotational}
is taken into account. A detailed description of our scheme is presented
in Supplementary Note 1.

In the classical limit, the rotational motion of rigid molecules driven
by external time-dependent fields was modeled with the help of Euler's
equations \citep{Goldstein2002Classical}
\begin{equation}
\mathbf{I}\bm{\dot{\Omega}}=(\mathbf{I}\bm{\Omega})\times\bm{\Omega}+\mathbf{T},\label{eq:Eulers-equations}
\end{equation}
where $\mathbf{I}=\mathrm{diag}(I_{a},I_{b},I_{c})$ is the moment
of inertia tensor $(I_{a}\leq I_{b}\leq I_{c})$, $\bm{\Omega}=(\Omega_{a},\Omega_{b},\Omega_{c})$
is the angular velocity, and $\mathbf{T}=(T_{a},T_{b},T_{c})$ is
the external torque expressed in the molecular frame. The latter is
given by $\mathbf{T}=\boldsymbol{\upmu}\times Q^{T}\mathbf{E}$, where
$\mathbf{E}$ is the external electric field defined in the laboratory
frame, and $Q(t)$ is a $4\times3$ matrix composed of the elements
of a quaternion \citep{Coutsias2004The,Kuipers1999Quaternions}. A
quaternion is defined as a quadruplet of real numbers, $q=(q_{0},q_{1},q_{2},q_{3})$
and it has a simple equation of motion $\dot{q}=q\Omega/2,$ where
$\Omega=(0,\bm{\Omega})$ is a pure quaternion and the quaternions
multiplication law is implied \citep{Coutsias2004The,Kuipers1999Quaternions}.
To simulate the behavior of a thermal ensemble, we use the Monte Carlo
approach. For each molecule, we numerically solve the system of Euler
{[}Eq. (\ref{eq:Eulers-equations}){]} and quaternion equations of
motion using the standard fourth order Runge-Kutta algorithm. Our
ensemble consists of $N=10^{7}$ sample molecules, which were initially
isotropically oriented. The initial uniform random quaternions were
generated using the recipe from Ref. \citep{Lavalle2006Planning}.
Initial molecular angular velocities were generated according to the
Boltzmann distribution $f(\Omega_{i})\propto\exp\left[-I_{i}\Omega_{i}^{2}/\left(2k_{B}T\right)\right],\;i=a,b,c$,
where $T$ is the rotational temperature and $k_{B}$ is the Boltzmann
constant.

\emph{Simulation results}.---In this work, we used methyl chloride
($\mathrm{CH_{3}Cl}$) and propylene oxide (PPO, $\mathrm{CH_{3}CHCH_{2}O}$)
molecules, as typical examples of symmetric- and asymmetric-top molecules,
respectively. Moments of inertia and components of the dipole moment
of these molecules are provided in Supplementary Note 2.

Figure \ref{fig:Quantum-classical-singlepulse} shows time-dependent
ensemble-averaged projection of the dipole moment, $\braket{\mu_{Z}}(t)$
on the laboratory $Z$ axis, along which the THz pulse is polarized.
The angle brackets $\braket{\cdot}$ denote thermal averaging over
the initial molecular state. The initial temperature is set to $T=5$
K. The two other averaged projections of the molecular dipole, $\braket{\mu_{X}}(t)$
and $\braket{\mu_{Y}}(t)$ are zero, due to the axial symmetry of
the system. On the short time scale (first $\approx10$ ps), the quantum
results are in good agreement with the classical ones. Both symmetric-
{[}Fig. \ref{fig:Quantum-classical-singlepulse}(a){]} and asymmetric-top
{[}Fig. \ref{fig:Quantum-classical-singlepulse}(b){]} systems immediately
respond to the THz pulse by a transient dipole signal. However, long
after the pulse, the orientation does not completely disappear and
a long-lasting persistent dipole signal remains (see insets of Fig.
\ref{fig:Quantum-classical-singlepulse}). This remarkable behaviour
is the main result of our work.

\begin{figure}[!t]
\centering{}\includegraphics[width=1\linewidth]{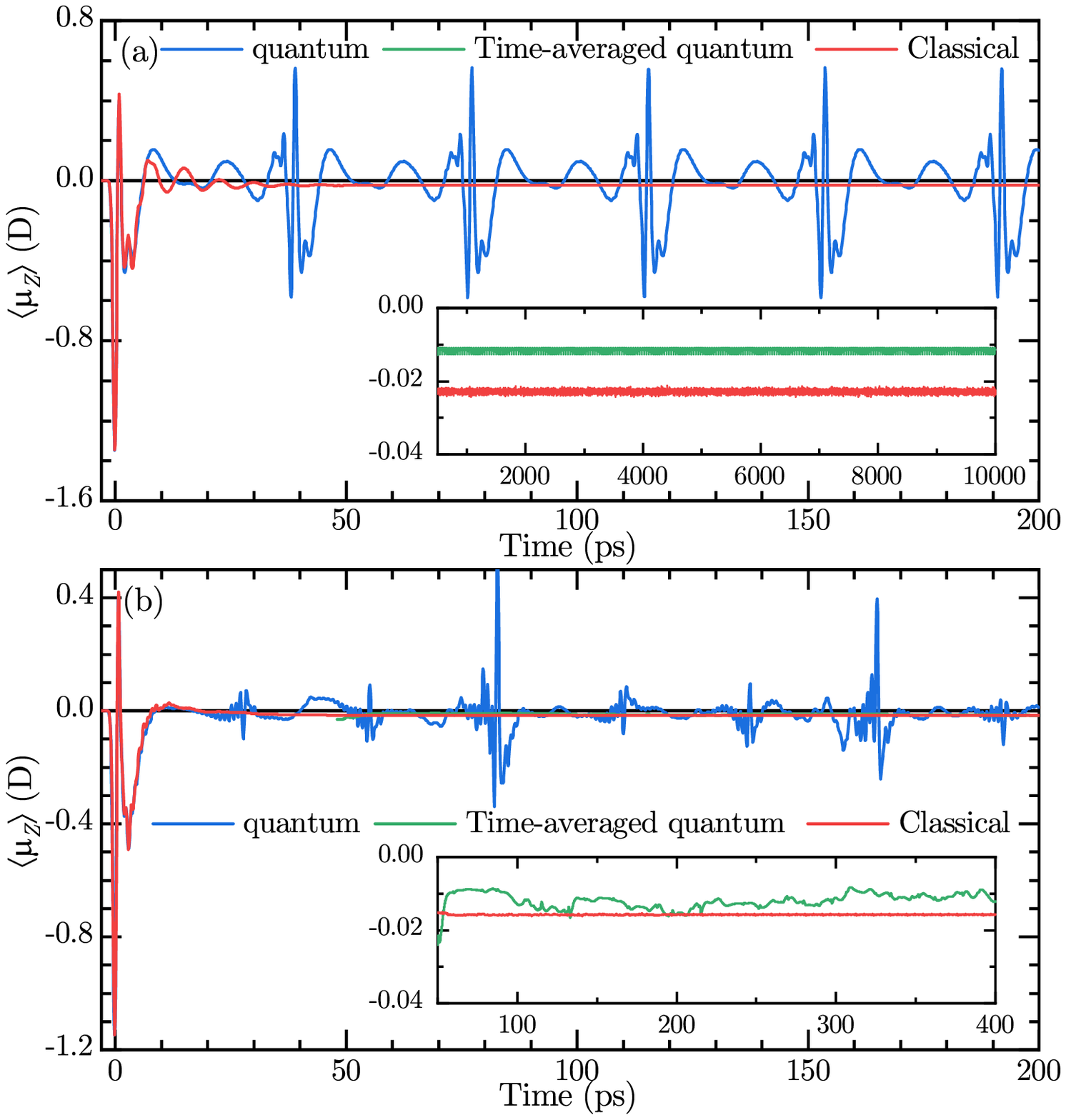} \caption{Ensemble averaged $Z$-projection of the dipole moment in the laboratory
frame, $\braket{\mu_{Z}}$ as a function of time for (a) $\mathrm{CH_{3}Cl}$
and (b) $\mathrm{PPO}$ molecules. The results of the quantum and
classical simulations are shown in blue and red lines, respectively.
The green curve represents the sliding time average defined by $\overline{\langle\mu_{Z}\rangle(t)}=(\Delta t)^{-1}\int_{t-\Delta t/2}^{t+\Delta t/2}\mathrm{d}t'\langle\mu_{Z}\rangle(t')$,
where (a) $\Delta t=1000$ ps and (b) $\Delta t=100$ ps, respectively.
The insets show amagnified portion of the signals. \label{fig:Quantum-classical-singlepulse}}
\end{figure}

\begin{figure*}[!t]
\centering{}\includegraphics[width=1\linewidth]{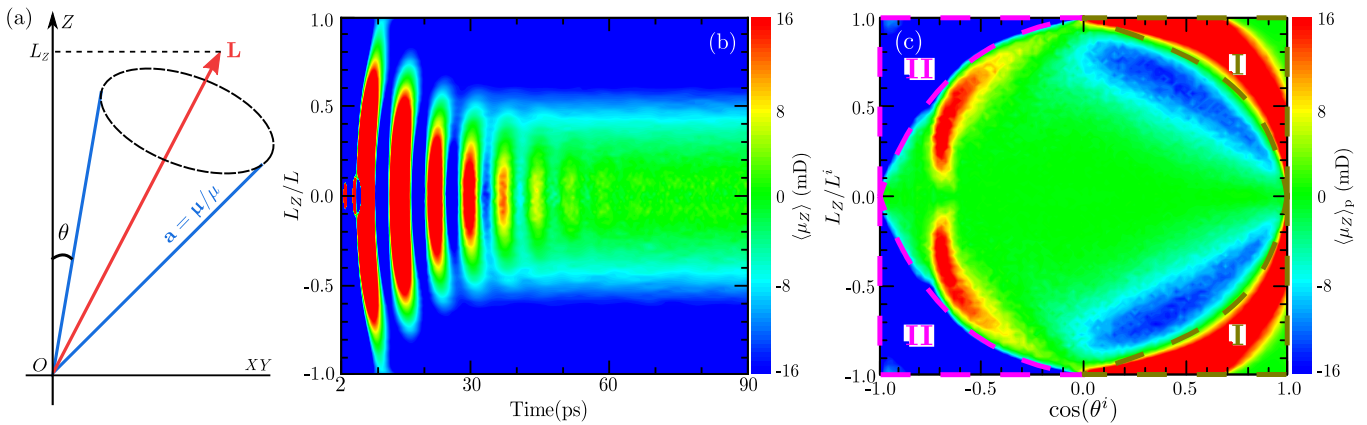}\caption{Thermal ensemble of classical symmetric-top molecules excited by $Z$-polarized
THz pulse. (a) Illustration of the precession motion of the molecular
symmetry axis $\mathbf{a}$ (in blue) about the angular momentum vector
$\mathbf{L}$ (in red), $\theta$ is the polar angle between $\mathbf{a}$
($\boldsymbol{\upmu}$) and the $Z$-axis. (b) The dipole signal $\braket{\mu_{Z}}$
as a function of time and $L_{Z}/L$. (c) Permanent value of the dipole
signal $\braket{\mu_{Z}}_{\mathrm{p}}$ after the pulse as a function
of the initial conditions before the pulse: $\cos\theta^{i}$ and
$L_{Z}/L^{i}$. Two regions, I and II, are marked in (c). The color
scales in (b) and (c) are in the units of millidebye (mD). \label{fig:explanation}}
\end{figure*}

In the classical limit, the long lasting orientation of symmetric-
and asymmetric-top molecules remains constant indefinitely (until
the collision destroy the orientation). Quantum simulations, on the
other hand, exhibit characteristic quasi periodic beats \citep{Felker1992Rotational}
(quantum revivals). However, a moving window average of the quantum
results (see captions of Fig. \ref{fig:Quantum-classical-singlepulse})
remains constant as the classically predicted orientation behavior.
As can be seen in caption of Fig. \ref{fig:Quantum-classical-singlepulse}(a),
the time-averaged quantum signal is constant in the case of symmetric-top
molecules. In the case of asymmetric-top molecules {[}Fig. \ref{fig:Quantum-classical-singlepulse}(b){]}
the time-averaged signal is long-lasting as well (thousands of ps
in the present case). On the even longer time-scale, the absolute
value of orientation slowly decreases and eventually changes its sign
(not shown). The difference between symmetric- and asymmetric-top
molecules stems from the fact that quantum-mechanical states of the
latter are non-degenerate and have different parity \citep{zare1988Angular}.
As a consequence, any state that is internally oriented at some point
in time cannot be an eigenstate and will oscillate between being oriented
and anti-oriented, an effect known as dynamical tunneling \citep{Keshavamurthy2011Dynamical}.

The results presented in Fig. \ref{fig:Quantum-classical-singlepulse}(a)
showing a series of identical revivals are valid within the rigid
rotor approximation. Two additional factors should be mentioned: centrifugal
distortion and radiation emission due to fast changes in the dipole
moment. The centrifugal distortion leads to eventual decay of the
periodic quantum revival peaks \citep{Babilotte2016Observation,Damari2017Coherent}
due to the dephasing of the rotational states. However, dipole signal
averaged over multiple revival periods is hardly affected (see Supplementary
Note 3). Fast variation of the dipole signal during each revival event
may lead to radiative emission and gradual reduction of the rotational
energy \citep{Babilotte2016Observation,Damari2017Coherent}. However,
in the case of a rarified molecular gas considered here, the estimated
relative energy loss during a single revival is negligible.


Recently, a related phenomenon of persistent orientation was reported \citep{Tutunnikov2019Laser, Tutunnikov2020Observation}, where twisted-polarization laser pulses were applied to the chiral molecules. Here, on the other hand, we use a single unshaped THz pulse, chirality is not required and the underlying mechanism is different, as is discussed in the next section.

\emph{Qualitative discussion}.---To understand the origin of the long
lasting orientation shown in Fig. \ref{fig:Quantum-classical-singlepulse},
we first analyze the case of an ensemble of classical symmetric-top
molecules excited by a $Z$ polarized THz pulse. The motion of free,
prolate for definiteness, symmetric-top is defined by a simple vectorial
differential equation $\dot{\mathbf{a}}=\left(\mathbf{L}/I\right)\times\mathbf{a},$
where $\mathbf{a}$ is a unit vector along the molecular axis of symmetry,
$I$ is the moment of inertia $(I_{a}<I_{b}=I_{c}\equiv I)$, and
$\mathbf{L}$ is the vector of angular momentum. Explicit solution
of this equation is provided in Supplementary Note 4. Figure \ref{fig:explanation}(a)
shows the general motion of a symmetric-top, namely precession of
the unit vector $\mathbf{a}$ about the vector $\mathbf{L}$ at rate
$L/I$, where $L$ is the magnitude of the angular momentum. The time-averaged
projection of the dipole on the $Z$ axis is given by
\begin{equation}
\overline{\mu_{Z}}=\mu\lim_{\tau\rightarrow\infty}\frac{1}{\tau}\int\limits _{0}^{\tau}\mathbf{a}(t)\cdot\mathbf{Z}\mathrm{d}t=\mu\frac{L_{Z}L_{a}}{L^{2}},
\end{equation}
where $\mu$ is the magnitude of the molecular dipole moment. The
solution $\mathbf{a}(t)$ describes the field-free stage of the motion,
therefore $L_{a}$ and $L$ in the above equation are taken at the
end of the THz pulse. Notice that the projection of the angular momentum
on the axis of the pulse, $L_{Z}$ is a constant of motion. The projection,
$L_{a}$ of the angular momentum on the molecular symmetry axis can
be expressed in terms of $L_{X}$, $L_{Y}$, $L_{Z}$, and the Euler
angles \citep{zare1988Angular}: $L_{a}=L_{X}\sin\theta\cos\phi+L_{Y}\sin\theta\sin\phi+L_{Z}\cos\theta.$

Thus, we have $\overline{\mu_{Z}}/\mu=(L_{Z}/L)^{2}\cos\theta+(L_{X}L_{Z}/L^{2})\sin\theta\cos\phi+(L_{Y}L_{Z}/L^{2})\sin\theta\sin\phi$.
Denoting $\cos\theta=x$ and $L_{Z}/L=y=\cos\theta_{1}$, the ensemble
averaged dipole moment is given by
\begin{equation}
\overline{\braket{\mu_{Z}}}\propto\mu\int\limits _{-1}^{1}xP_{0}(x)\mathrm{d}x,\label{eq:integral}
\end{equation}
where $P_{0}\propto\int y^{2}L^{2}P_{1}(x,y,L)\mathrm{d}L\mathrm{d}y$
and $P_{1}(x,y,L)$ is the joint probability distribution of $\cos\theta$,
$\cos\theta_{1}$ and $L$ at the end of the THz pulse. Terms proportional
to $\sin\phi$ and $\cos\phi$ in the expression for $\overline{\mu_{Z}}/\mu$
do not contribute to the average, because a THz pulse does not affect
a uniform distribution of the angle $\phi$. For initial temperature
$T=0$ K, $\overline{\mu_{Z}}$ vanishes because $L_{Z}\equiv0$ before
and after the THz pulse. Finally, in the limit of a pulse of vanishing
duration ($\kappa\rightarrow\infty$, see Fig. \ref{fig:THz-field}),
the permanent orientation tends to zero because $P_{0}$ keeps the
initial symmetric distribution just after the pulse. As an illustration,
the joint probability distribution $P_{1}$ can be written down explicitly
for a model system of thermalized symmetric-top molecules ($I/I_{a}=w>1$)
subject to a constant dc field of amplitude $E_{0}$ which is then
abruptly switched off. In this case, $P_{1}$ is separable and is
given by $P_{1}(x,y,l)\propto f(x)g(y,l)=\exp\left(\epsilon x\right)\exp\left\{ -l^{2}\left[1+y^{2}\left(w-1\right)\right]\right\} $,
where $\epsilon=\mu E_{0}/k_{B}T$ and $l=L/\sqrt{2Ik_{B}T}$ (see
Supplementary Note 5). Clearly, $\overline{\braket{\mu_{Z}}}\neq0$,
because $\exp\left(\epsilon x\right)$ is not symmetric in the interval
$x\in[-1,1]$ for $\epsilon>0$.

To explore the long lasting orientation and to better understand its
physical origin, we consider the underlying dynamics in more detail.
After the end of the pulse, at $t>2$ ps (see Fig. \ref{fig:THz-field}),
the ratio $L_{Z}/L$ is constant. Figure \ref{fig:explanation}(b)
shows the ensemble averaged dipole moment, $\braket{\mu_{Z}}$ in
a three dimensional color-coded plot, as a function of time and the
(constant) ratio $L_{Z}/L$. As can be seen in the figure, right after
the pulse ($t=2$ ps) approximately all the molecules have a negative
$Z$ projection of their dipoles. Moreover, Fig. \ref{fig:explanation}(b)
shows that, at long times, molecules with $L_{Z}\simeq\pm L$ (upper
and lower blue bands) contribute to the overall negative dipole signal,
while for molecules with $|L_{Z}/L|\ll1$ (central portion of the
figure), the time-averaged dipole signal tends to zero. In the absence
of external fields, the molecules with angular momentum along $\pm Z$
axis, and the symmetry axis initially directed along $-Z$, will continue
to precess around the $-Z$ direction forever. In contrast, the contribution
of molecules with angular momentum lying in the $XY$ plane ($|L_{Z}/L|\ll1$)
averages to zero and they do not contribute to the long lasting orientation.
In the special case of initial temperature of $T=0$ K, $L_{Z}$ remains
identically zero after the pulse ($L_{Z}$ is conserved), thus forbidding
the permanent orientation.

The physical origin of the permanent (negative) orientation observed
in Fig. \ref{fig:explanation}(b) can be further analysed by considering
it as a function of \emph{the initial} orientation factor, $\cos\theta^{i}$
and the value of $L_{Z}/L^{i}$ before the pulse. Figure \ref{fig:explanation}(c)
shows the value of $\langle\mu_{Z}\rangle$ taken at times after 100
ps {[}see Fig. \ref{fig:Quantum-classical-singlepulse}(a){]}, denoted
as $\langle\mu_{Z}\rangle_{\mathrm{p}}$, as a function of the initial
conditions. As a reminder, $\theta^{i}$ is the initial angle between
the dipole moment and the $Z$-axis {[}see Fig. \ref{fig:explanation}(a){]}.
The figure shows that molecules with $|\cos\theta|\simeq1$ and $L_{Z}/L^{i}=\pm1$
(regions denoted by I and II, blue and red in the figure) maintain
the nonzero permanent orientation, while the orientation of molecules
in other regions tends to zero. To understand the reason for this,
we consider molecules in each region separately and follow their time
evolution in the presence of the THz pulse.

Initially, molecules in region I precess about the $+Z$ direction,
because $L_{Z}\simeq\pm L^{i}$ and $\cos\theta^{i}\simeq1$. During
the first phase of the THz pulse, when the field points along $-Z$
(see Fig. \ref{fig:THz-field}), it tends to flip the molecular dipoles.
In contrast, during the second phase, when the field is along $+Z$
(see Fig. \ref{fig:THz-field}), it pushes the dipoles towards $+Z$,
thus facilitating the permanent orientation. Molecules in region II,
which initially precess about $-Z$ ($L_{Z}\simeq\pm L^{i}$, $\cos\theta^{i}\simeq-1$)
respond oppositely as compared to the molecules in region I. For the
chosen parameters of the THz pulse (see Fig. \ref{fig:THz-field}),
there is an imbalance between these two groups resulting in the excess
of molecules precessing about $-Z$ axis. This leads to the overall
negative permanent orientation.

\begin{figure}[!t]
\centering{}\includegraphics[width=1\linewidth]{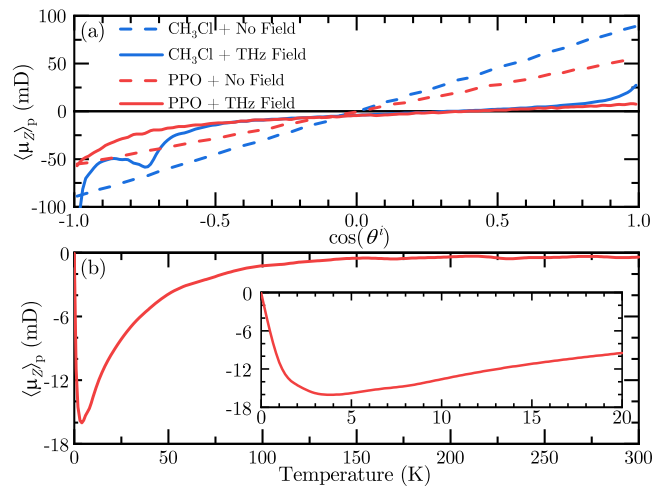}\caption{(a) Classically calculated permanent values of the dipole signals
$\langle\mu_{Z}\rangle_{\mathrm{p}}$ as a function of initial angular
position $\cos\theta^{i}$ for four cases: $\mathrm{CH_{3}Cl}$ without
field (dashed blue), $\mathrm{CH_{3}Cl}$ and THz field (solid blue),
$\mathrm{PPO}$ without field (dashed red), and $\mathrm{PPO}$ and
THz field (solid red). (b) Classically calculated permanent values
of the dipole signal $\langle\mu_{Z}\rangle_{\mathrm{p}}$ as a function
of temperature for $\mathrm{PPO}$ molecule. \label{fig:Temperature}}
\end{figure}

The qualitative mechanism described above for the case of symmetric-top
molecules is essentially the same for asymmetric-top molecules as
well. In order to establish the similarity between the symmetric-
and asymmetric-top molecules, Fig. \ref{fig:Temperature}(a) depicts
the permanent values of the dipole signals $\langle\mu_{Z}\rangle_{\mathrm{p}}$
as a function of initial orientation quantified by $\cos\theta^{i}$
for the previously considered methyl chloride and propylene oxide
molecules. In both cases, the THz pulse preferentially supports the
molecular precession about the $-Z$ direction, as opposed to $+Z$,
so that the overall long-time dipole value is negative.

It was pointed out earlier that at $T=0$ K permanent orientation
is impossible. Also, as $T\rightarrow\infty$ the orientation disappears,
as well. This non-monotonous temperature dependence leads to the existence
of an optimal initial temperature that will provide the maximal (in
absolute value) permanent orientation. Indeed, Fig. \ref{fig:Temperature}(b)
depicts the temperature dependence of the permanent dipole signal,
calculated classically. As can be seen in the figure, for the pulse
parameters used here, the maximal persistent orientation is achieved
for initial temperature of $T\approx4$ K.

\emph{Summary}.---We have theoretically demonstrated a new phenomenon
of persistent orientation of symmetric- and asymmetric-top molecules
excited by a single THz pulse. The orientation was shown to persist
long after the end of the pulse for both types of molecules. Analysis
of the relatively simple special case of symmetric-top molecules reveals
the underlying classical mechanism and provides a detailed understanding.
The mechanism is general and relies on the symmetry breaking of the molecular ensemble caused by the interaction with the external field.
Similar mechanism may lead to persistent orientation effect in the cases
of symmetric- and asymmetric-tops excited by two-color laser fields
\citep{De2009Field,Oda2010All,Wu2010Field,Frumker2012Oriented,Lin2018All}.
The fast transient dipole signal and its long-time persistent
component may be directly measured with the help of second (or higher
order) harmonic generation, which is sensitive to the lack of inversion
symmetry \citep{Frumker2012Oriented}.
\\

This work was supported by the Israel Science Foundation (Grant No.
746/15), the ISF-NSFC joint research program (Grant No. 2520/17).
I.A. acknowledges support as the Patricia Elman Bildner Professorial
Chair. This research was made possible in part by the historic generosity
of the Harold Perlman Family.

\end{document}